Improved liver T1rho measurement precision with a breathhold black blood single shot fast spin echo acquisition: a validation study in healthy volunteers.

Yì Xiáng J Wáng *MMed PhD* [1], Min Deng *MMed*[1], Gladys G Lo *MD*[2], Queenie Chan *PhD*[3], Yuan Jing *PhD*[4], Weitian Chen *PhD*[1],

[1]Department of Imaging and Interventional Radiology, The Chinese University of Hong Kong, Prince of Wales Hospital, Shatin, New Territories, Hong Kong SAR
[2]Department of Diagnostic & Interventional Radiology, Hong Kong Sanatorium & Hospital, Happy Valley, Hong Kong SAR
[3] MR Clinical Science, Philips Healthcare Greater China, Hong Kong SAR
[4] Medical Physics and Research Department, Hong Kong Sanatorium & Hospital, Happy Valley, Hong Kong SAR

Running title: a breathhold black blood single shot T1rho acquisition

*Funding:* This study was partially supported by grants from the Research Grants Council of the Hong Kong SAR (Project No. 476313 and Project No. SEG CUHK02).
*Conflict-of-interest:* Queenie Chan is an employee of Philips Healthcare. The other authors declare t no conflict of interest.

Correspondence to: Yì-Xiáng Wáng; e-mail: yixiang_wang@cuhk.edu.hk

Abstract

*Purpose*: To explore the usability and normal T1rho value of liver parenchyma with a novel single breathhold black blood single shot fast spin echo acquisition based liver imaging sequence.

*Materials and Methods*: The institutional Ethics Committee approved this study. In total 19 health subjects (10 males, 9 females; mean age: 37.4 yrs; range: 23-54 yrs) participated in the study. 11 subjects had liver scanned twice in the same session to access scan-rescan repeatability. 12 subjects had liver scanned twice in two sessions with 7-10 days' interval to access scan-rescan reproducibility. MR was performed with a 3.0 T scanner with dual transmitter. The MR sequence allows simultaneous acquisition of 4 spin lock times (TSLs: 0msec, 10 msec, 30 msec, 50msec) in 10 second, and the spin-lock frequency was 500Hz. Inherent black blood effect of fast spin echo and double inversion recovery were utilized to achieve blood signal suppression. Three axial slices were obtained with a resolution of $1.5 \times 1.5 \times 6.00$ mm3. Region-of-interest method was used to measure liver T1 rho value.

*Results*: The technique demonstrated good image quality and minimal artifacts. For liver parenchyma, Bland-Altman plot showed the scan-rescan repeatability mean difference was 0.025 msec (95% limits of agreement: -1.163 to 1.213 msec), and intraclass correlation coefficient (ICC) was 0.977. The scan-rescan reproducibility mean difference was -0.075 msec (95% limits of agreement: -3.280 to 3.310 msec), and ICC was 0.820 which is better than the ICC of 0.764 of a previous bright blood multi-breath hold gradient echo acquisition technique. The liver T1rho value was $39.9 \pm 2.4$ msec (range: 36.1 - 44.2 msec), which is lower than the value of $42.8 \pm 2.1$ msec acquired with the previous bright blood technique.

*Conclusion*: This study validated the application of a single breathhold black blood single shot fast spin echo acquisition based for human liver T1rho imaging. The lower liver parenchyma T1rho value and higher scan rescan reproducibility may improve of the sensitivity of this technique.

Introduction

Chronic liver disease is a major public health problem worldwide. Hepatitis virus is the most common blood-borne infection in USA and worldwide [1,2]. Chronic viral hepatitis can lead to hepatic fibrosis, cirrhosis and hepatocellular carcinoma and is the leading cause of liver transplantation nationwide [3]. The epidemic trend of chronic liver disease is expected to increase owing to an aging population, the growing epidemic of obesity and non-alcoholic steatohepatitis. To date, noninvasive diagnostic tests available from clinical practice are not sensitive or specific enough to detect liver injury at early stages [4]. Liver biopsy remains the standard of reference for the diagnosis and staging of liver fibrosis. However, it is an invasive procedure with potential complications [5]. Histologic assessment of fibrosis is also an inherently subjective process and subject to sampling variability [6]. A noninvasive and quantitative technique for assessing liver fibrosis and monitoring disease progression or therapeutic intervention is highly desirable.

Compared with molecular imaging based on injectable agents, MR based molecular imaging techniques have the major advantage that the regulatory hurdle is much less [6-18]. T1rho (T1$\rho$) is the relaxation rate constant of transverse magnetization for the duration of the spin-lock radiofrequency (RF) pulse. In magnetic resonance T1rho imaging, the equilibrium magnetization ($M_0$) is rotated into the transverse plane first. This magnetization in the transverse plane relaxes as normal free induction decay in the presence of a weaker on-resonant continuous wave radiofrequency called spin-lock pulse. It is sensitive to both static processes and low-frequency motional processes, so it can be used to detect macromolecular compositions and proton exchange in various tissues [19, 20]. Wang *et al* and Zhao *et al* described the usefulness of T1rho MR imaging for assessment in rat liver fibrosis models, suggesting the increase T1rho value is associated with collagen deposition in the liver fibrosis models but less so with inflammatory edema [21,22]. Studies in human subjects, both in healthy volunteers and liver fibrosis patients, were soon followed [23-31]. With a two-dimensional fast-field echo sequence, Deng *et al* and Zhao *et al* reported the mean T1rho value to be 42.8±2.1 msec ($B_0$=3T, $B_1$=500Hz) in healthy volunteers [23-25]. This value was further confirmed by Allkemper *et al* [27] with a respiratory-gated 3D fast-field echo sequence [$B_0$=1.5T, $B_1$=500Hz]. Allkemper *et al* further reported that liver T1rho values did not correlate with necroinflammatory activity, degree of steatosis or presence of iron load, but it was significantly associated with the Child-Pugh staging of the patients. In patients with chronic liver diseases, Takayama *et al* [30] demonstrated liver T1rho values showed significant positive correlations with the serum levels of total bilirubin, direct bilirubin, and indocyanine green (ICG-R15) , and significant negative correlations with the serum levels of albumin and γ-glutamyl transpeptidase.

Most of these previous reported methods were implemented with multiple breath-holds or respiratory gating, which is difficult to prevent small motion displacement between images acquired at different spin lock duration times (TSLs). These studies were also based on fast gradient echo acquisitions without suppression of blood signal, while blood vessels may contribute to elevating T1rho measurement for normal parenchyma [27, 29]. The presence of bright blood signal also increases sensitivity of T1rho quantification to motion; while as discussed in literatures [29], even minor spatial misregistration between images acquired with different TSL can lead to artificially elevated T1rho measurement for liver parenchyma. More recently Singh *et al* [31] proposed a pulse sequence enabling acquisition of single slice 2D T1rho MR imaging with four different TSLs in a single breath-hold of 12 second, and a small cohort study demonstrated the potential of detecting early stage liver fibrosis in patients. However, the longest TSL used in that study was 30 ms; therefore lower than the expected normal liver tissue T1rho value. Yang e*t al* [32] reported the choice of MR sequences can impact the T1rho value of normal liver parenchyma. Additionally, despite the difference of the mean liver T1rho values between the healthy volunteers and cirrhosis patients was statistically significant for the sequences they tested, namely, multi-slice and 3-dimentional sequence with a block RF pulse used for spin locking or a stretched type adiabatic pulse used spin-locking; there was substantial over-lap with higher-end value of healthy livers and lower-end value of cirrhotic livers.

Chen *et al* recently proposed a 2D black blood T1rho MR Relaxometry technique for liver imaging [33]. Black blood effect of 2D Fast Spin Echo (FSE) sequence and double inversion recovery (DIR) were utilized to achieve blood signal suppression. The current study describes the application of this technique on healthy volunteers. With reference to previously acquired data using a 2D bright blood gradient echo based sequence [23, 24], the following aspects were evaluated, 1) tolerability (i.e. comfortability) of this sequence for volunteers, 2) imaging quality and potential artifacts of images acquired with this sequence, 3) imaging repeatability and reproducibility of this sequence, and 4) the range and mean of normal liver parenchyma T1rho value.

Materials and Methods

The volunteer study was conducted with the approval of the Institutional ethics committee. The volunteers provided informed consent. In total 19 healthy subjects (10 males, 9 females; mean age: 37.4 yrs; range: 23-54 yrs) participated in the study. 12 subjects had liver T1rho scanned twice in the same session to assess scan-rescan repeatability. 12 subjects had T1rho scanned twice in two sessions with 7-10 days' interval to assess scan-rescan reproducibility (table 1). The selection of subjects to undergo repeated scans was chosen based on the availability of the subjects to participate. Data were collected from a Philips Achieva TX 3.0 T scanner equipped with dual transmit (Philips Healthcare, Best, the Netherlands). A 32 channel cardiac coil (Invivo Corp, Gainesville, USA) was used as receiver and body coil was used as transmit. RF shimming was applied to reduce B1 inhomogeneity. The subjects were scanned in supine position. 2D axial images were acquired with phase encoding along anterior-posterior direction.

The pulse sequence scheme used in this study has been described in [33]. A RF pulse cluster that can achieve simultaneous compensation of B1 RF and B0 field inhomogeneity was used for T1rho preparation. The combination of double inversion recovery (DIR) and 2D Fast spin Echo (2DFSE) was used to achieve black blood effect [34, 35]. The parameters for MR imaging included: TR/TE 2,500/15 ms, in-plane resolution 1.5 mm × 1.5 mm, slice thickness 6 mm, SENSE acceleration factor 2, half scan factor (partial Fourier) 0.6, number of signal averaging 1, delay time for SPAIR 250 ms, delay time for DIR 720 ms, spin-lock frequency 500Hz. Images with four spin-lock times 01, 10, 30, 50msec were acquired [24, 36]. The entire single slice T1rho data sets with the four spinlock time were acquired within a single breathhold of 10 seconds. Three slices were acquired for each examination.

Breath-hold was trained for the volunteers before the scan started. We found that it is more likely that the diaphragm and liver position would shift if we ask the volunteers to hold their breath after full end-inspiration or full end-expiration, we chose therefore to ask volunteer to hold their breath during usual-depth breathing. In addition, a time delay was allowed between the scan operator to give 'hold-breath' instruction and to push the MR data acquisition start button, so that the volunteers had time to react to the 'hold-breath' instruction. The respiration-gating balloon was placed on the top of the volunteers' upper abdomen, and the quality of the 'hold-breath' was monitored on the respiration-triggering screen on the MR console.

All images were processed using Matlab R2015a (Mathworks, USA). T1rho maps were computed by using a mono-exponential decay model, as described by the following equation: $M\,(TSL) = A \cdot \exp\,(-\,TSL/T1\,rho)$, where $A$ is a constant scaling factor and TSL is the time of spin-lock. Non-linear least square fit with the Levenberg-Marquardt algorithm was applied. Maps of coefficient of determination (R2) were also generated for the evaluation of goodness of fit. Only T1rho values for pixels associated with $R^2 > 0.80$ were included in the subsequent region of interest (ROI) placement and T1rho analysis to eliminate the unreliable poorly fitted T1rho values due to artefacts [23]. With reference to T1rho image at TSL of 01msec and T1rho map, region-of-interest (ROI) measurement of liver T1 rho value were performed. Four to six ROIs were placed on each axial section of the liver parenchymal region, avoiding potential artifacts as well as potential high value residual vessel signal. A total of approximately 15 ROIs were obtained for each liver, and the mean value of these 15 ROIs was regarded as the value of the liver's T1rho. For reference, T1rho value of bilateral erector spinae muscle was also measured, with one ROI placed on each side.  This ROI based approach has been well established in our laboratory, with an ICC for inter-reader measurement reproducibility of 0.955 [23]. Additionally, our previous study demonstrated there was no difference in ROI-based approach and whole liver histogram-based approach [25]. Therefore the intra- or inter- reader reproducibility was not further investigated in this study.

For repeatability, repeated scans in the first exanimation were used for Bland-Altman plot and intraclass correlation coefficient (ICC) analysis (n= 11 subjects). For reproducibility, there were four subjects being scanned three times; the first scan and the second scan were used for Bland-Altman plot and ICC analysis (n= 12 subjects). According to Fleiss [36], an ICC value of <0.4 represents poor agreement, a value of >0.75 represents good agreement and a value between 0.4 and 0.75 represents fair to moderate agreement.

The results obtained from this study were compared with previously acquired liver T1rho MR data using a 2D bright blood gradient echo based sequence [23, 24]. For this purpose, T1rho map image quality comparison was performed on the T1rho maps from this study and our previous studies [23, 24, raw data available upon request]. Mean scores for all slices in an examination were used as the visual score for T1rho maps, being categorized as: 1), poor, 2), fair, 3), good and 4), excellent [32]. Artifacts in the images were recorded as: 1): non-existing (no artifact exist on T1rho map), 2) very mild (small areas of mild artifacts and does not affect T1rho quantification), 3) acceptable (artifacts apparently notable, but not in the liver, or sufficient area of liver parenchyma were artifact-free, and artifact can be avoided by placing suitable ROIs), and 4) non-acceptable artifacts for liver T1rho quantification. These scorings were achieved by consensus reading of two readers.

Results

All subjects were able to collaborate with scans and the 10sec breath-hold very well. Only one subject' initial image showed motion artifacts and the artifacts were recognised during the examination and consequentially repeat scan was successfully obtained. Typical T1rho MR Images of healthy volunteer normal liver are shown in Fig 1. It can be seen fat signal and large vessel are satisfactorily suppressed. Single breath-hold T1rho mapping technique demonstrated good image quality and minimal artifacts Fig 2. The visual image quality score was 2.58±0.46 and 2.04±0.33 for the new sequence and previous sequence [23, 24] respectively (Fig 2A, p=0.002), and artifacts grades was 2.44±0.43 and 3.00±0.300 for the new sequence and previous sequence [23, 24] respectively (Fig 2B, p=0.001). Only

For liver parenchyma, Bland-Altman plot showed the scan-rescan repeatability mean difference was 0.025 msec with 95% limits of agreement of -1.163msec to 1.213 msec [Fig 3A], the associated ICC was 0.977. The scan-rescan reproducibility mean difference was -0.075 msec with 95% limits of agreement of -3.280 msec to 3.310 msec [Fig 3B], the associated ICC was 0.820. Thus, the scan-rescan reproducibility ICC in study is better than the ICC of 0.764 previously reported [23]. The physiological liver T1rho value was 39.9 ± 2.4 msec (range: 36.1-44.2 msec), which is consistent with expectation as bright blood techniques had reported measured liver T1rho value being around 42.8±2.1 msec [23, 24].

The erector spinae muscle T1rho was 32.0±1.7 msec (range: 28.5-36.2 msec, supplementary table 1). The association between liver parenchyma T1rho measurement and back muscle T1rho measurement was not significant (Fig 4, p=0.763).

Discussion

Both animal experiments and clinical studies have revealed that liver fibrosis, even early cirrhosis, is reversible, treatment with combined therapies on underline etiology and fibrosis simultaneously might expedite the regression of liver fibrosis and promote liver regeneration [39-40]. Earlier stage liver fibrosis is more amenable to therapeutic intervention. Even when the underline etiology of liver fibrosis could not be eradicated, therapies on liver fibrosis might help restrict the disease progression to cirrhosis [39]. Therefore early detection of liver fibrosis and treatment monitoring is of paramount importance.

In this study we studied a pulse sequence for quantitative T1rho imaging of liver based on single shot fast/turbo spin echo (SSFSE/SSTSE) acquisition [33]. The sequence in this study was implemented such that data sets with 4 TSL can be acquired within one single scan, which reduces its susceptibility to motion displacement occurred at different TSL [27, 29, 33]. Fast spin echo acquisition in combination with double inversion recovery allows blood suppression [33, 41, 42]. T1rho-prepapration is susceptible to the off-resonance effect, and therefore it is desirable to suppress fat signal. SPectral Attenuated Inversion Recovery (SPAIR) was used in the pulse sequence which applies a 180 degree adiabatic inversion pulse tune to the fat resonance. The time delay of SPAIR was chosen such that fat signal is zero at the beginning of T1rho-prepapration. Recently Singh *et al*. [31] reported a single slice acquisition with a breath-hold of 12 seconds at 1.5 T. However, the maximum TSL is limited to 30 ms in their study. Higher field strength can impose further restriction on the maximum duration of the spinlock radiofrequency pulse. However, parallel transmit provides reduced specific absorption rate (SAR) compared to conventional transmit. Our scanner is configured with dual-transmit and the SAR of our pulse sequence is within the US Food and Drug Administration limit under this configuration.

SSFSE is a standard clinical sequence for anatomical assessment of liver. It is feasible to tune the refocusing flip angle and the echo time in the proposed pulse sequence without altering mono-exponential relaxation model for T1rho quantification [33]. Desired image contrast can be obtained by tuning refocusing flip angle and echo time. Therefore, the proposed pulse sequence potentially can be used for simultaneous anatomical and biochemical (T1rho) assessment of liver, which may facilitate its incorporation into clinical protocol.

In this study, all 19 volunteers tolerate the breath hold well (=10 sec), and good quality T1rho maps were obtained with vessel signal well suppressed. Good scan-rescan reproducibility and repeatability was demonstrated. Despite 6 spin lock times was used in our previous studies [23, 24], the new sequence used in this study provided better T1rho map image quality, less artifacts, and higher scan rescan reproducibility (such as ICC improved from 0.764 to 0.820), and scan rescan repeatability in the same session was as high as 0.977. Note the visual image quality score of 2.58±0.46 and artifacts grades of 2.44±0.43 for T1rho reflect the high standard we placed for image quality assessment. The artifacts in the images acquired with the new sequence were mostly due to the pulsing of the aorta, and did not affect T1rho measurement of liver (Fig 1). The physiological T1rho value of liver parenchyma was measured to be 39.9 ± 2.4 msec in this study, which was lower than the value (42.8±2.1msec) obtained in reference 23 and 24 where no black-blood technique was applied. This new sequence mitigates potential quantification errors in liver due to motion in previous methods and therefore improves the robustness of T1rho imaging of liver. It is expected that lower physiological T1rho value measurement will increase the diagnostic confidence for liver fibrosis evaluation. In addition, the preliminary results of no association between liver parenchyma T1rho value and back muscle T1rho value may suggest there were no systemic upper-wards or down-wards measurements observed in this study.

To further improve this black blood T1rho relaxometry sequence, additional techniques can be incorporated, including more robust measurement [43-46] and faster data sampling and reconstruction [47, 48]. These techniques are being explored in our laboratory. Along with technical improvements, integrated liver multi-parametric imaging will likely improve the sensitivity and specificity for early stage liver fibrosis characterization, or a combination of MR readout and serum fibrosis markers and the use of decision trees, such as Bayesian prediction, will allow high validity for a diagnostic approach.

In conclusion, this study validated the application of black blood T1rho imaging in healthy subjects. The black blood effect together with a multiple spin lock times in a single breathhold acquisition helps to mitigate potential quantification errors in liver due to motion in previous methods, and therefore improves the measurement precision of T1rho imaging of liver with increased scan rescan reproducibility.

Fig legends.

Fig 1.  Typical T1rho MR Images of two healthy subjects (A, B) at the four spin lock time and T1rho maps with or without $R^2$ (>0.80) evaluation.

Fig 2. A comparison of T1rho map image quality (A) and artifacts (B) of black blood single shot fast spin echo acquisition (references 33) vs. bright blood multi-breath hold gradient echo acquisition (references 23, 24)

Fig 3. Bland-Altman Plots of liver parenchyma scan-rescan repeatability in the same session (A), and scan-rescan reproducibility when the examinations were performed with 7-10 days' interval (B).

Fig 4. Liver parenchyma T1ho value vs. erector spinae muscle T1ho value of individual volunteers.      No      significant      association      was      found.

Table 1, Liver parenchyma T1rho value (millisecond) in healthy volunteers.

| Case no. | scan_1.1 | scan_1.2 | scan_2.1 | scan_2.2 | scan_3.1 | scan_3.2 |
|----------|----------|----------|----------|----------|----------|----------|
| 1 | 44.2 | 44.1 | 44.6 | | | |
| 2 | 40.9 | 40.2 | | | | |
| 3 | 39.9 | 39.6 | 43.5 | | | |
| 4 | 42.2 | 41.5 | | | | |
| 5 | 40.2 | 39.8 | 36.9 | 36.9 | | |
| 6 | 37.4 | 37.3 | | | | |
| 7 | 36.2 | 36.5 | | | | |
| 8 | 40.9 | 41.5 | | | | |
| 9 | 39.8 | 38.9 | | | | |
| 10 | 40.2 | | 40.7 | | 42.9 | 42.9 |
| 11 | 38.1 | | 38.3 | | 38.5 | |
| 12 | 42.8 | | 43.9 | | | |
| 13 | 37.5 | | 36.4 | | | |
| 14 | 40.2 | | 41.5 | | | |
| 15 | 41.2 | | 41 | | | |
| 16 | 40.6 | | 39.8 | | | |
| 17 | 43.9 | | 43.3 | | 43.2 | |
| 18 | 37.1 | | 36.9 | | 37.4 | |
| 19 | 36.1 | 36.4 | | | | |

Note: Scan 1.1 vs. scan 1.2 comparison means scan re-rescan where two scans were conducted in the same secession and the subjects did not move in positioning, but sampled slice would be different. Scan 1.1 vs. scan 2.1 means scan re-rescan where two scans were conducted in two secessions with an interval of 7-10 days apart.

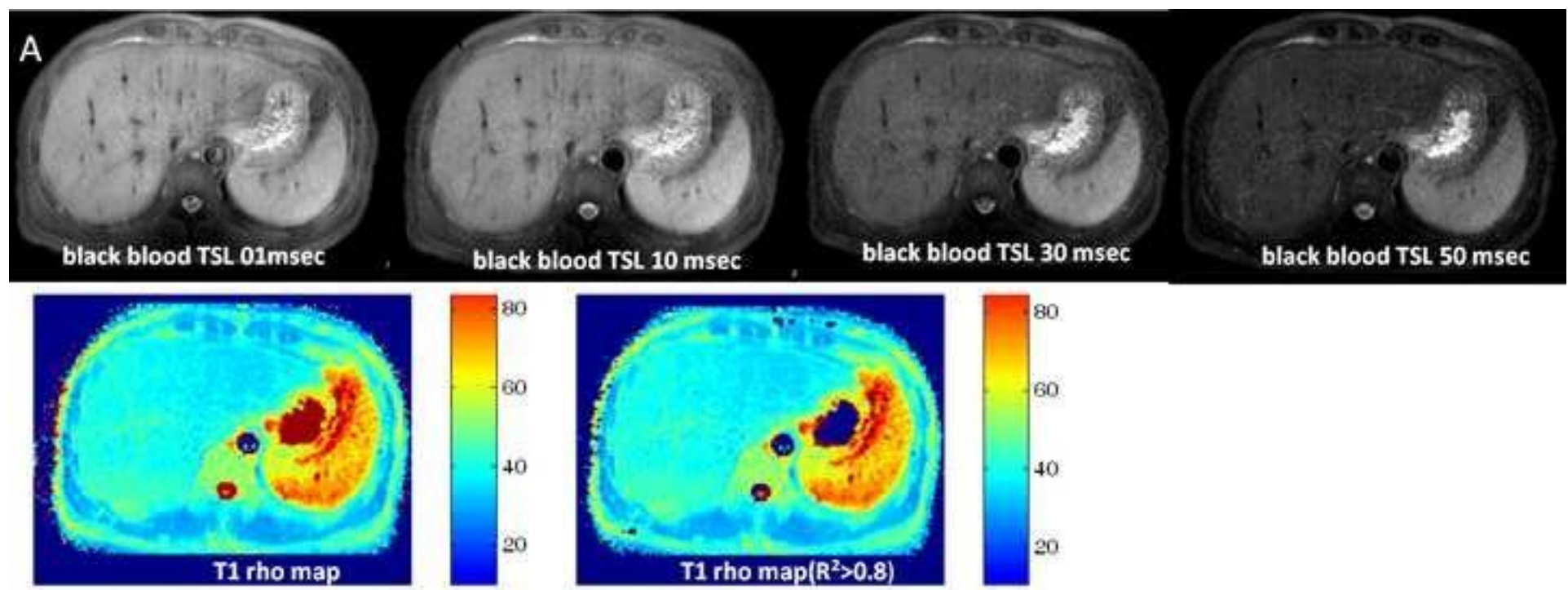

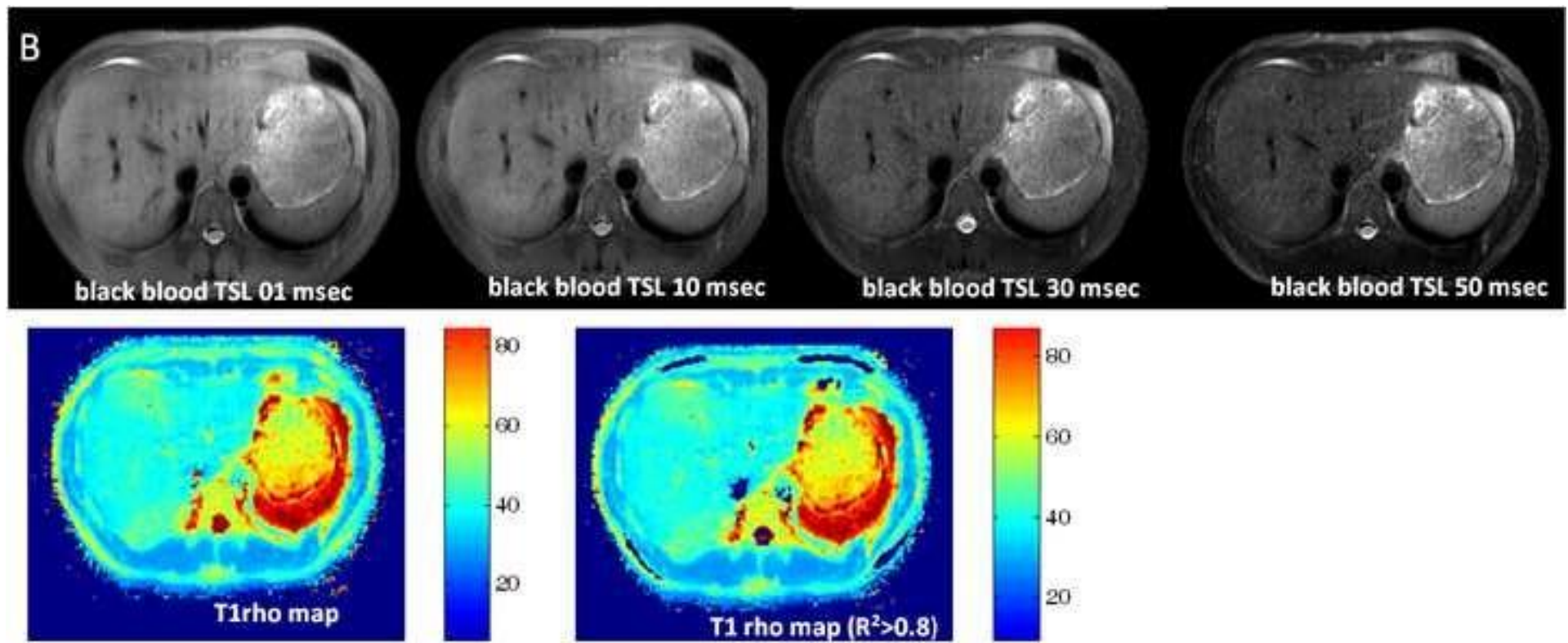

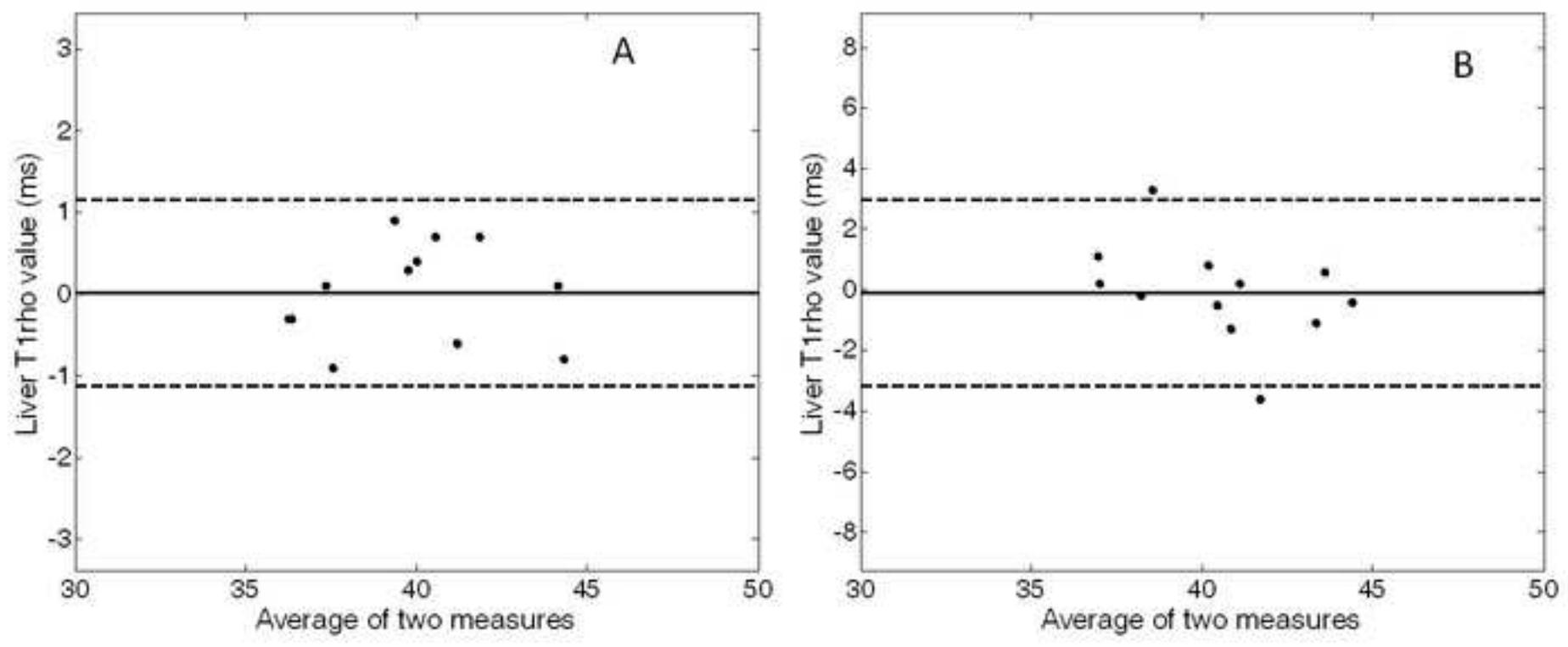

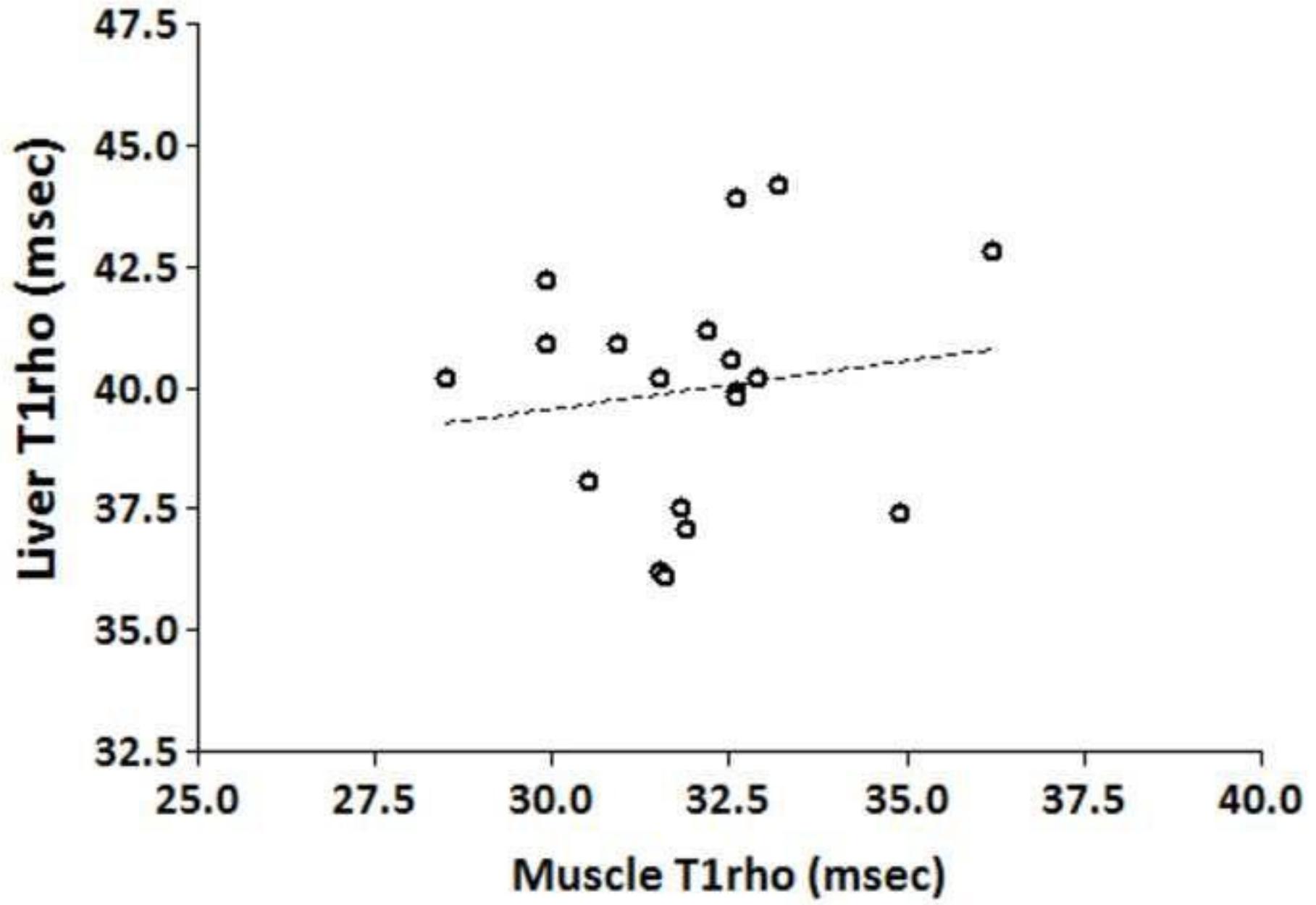